\documentclass[twocolumn,aas_macros]{article}
\usepackage{graphicx}
\usepackage{amsmath}
\usepackage{txfonts}
\usepackage{natbib}
\usepackage{fancyhdr}
\usepackage[affil-it]{authblk}
\usepackage[margin=0.4in]{geometry}
\usepackage{textcomp}
\usepackage{aas_macros}

\begin{document}

\lhead{Frequency-dependent time lags in Sgr A*}
\rhead{C.D.Brinkerink et al.}

\title{ALMA and VLA measurements of frequency-dependent time lags in Sagittarius A*: evidence for a relativistic outflow}

\author{Christiaan D. Brinkerink$^1$, Heino Falcke$^1$, Casey J. Law$^2$, Denis Barkats$^3$, Geoffrey C. Bower$^4$, Andreas Brunthaler$^5$, Charles Gammie$^6$, C. M. Violette Impellizzeri$^7$, Sera Markoff$^8$, Karl M. Menten$^5$, Monika Moscibrodzka$^1$, Alison Peck$^7$, Anthony P. Rushton$^{9,10}$, Reinhold Schaaf$^11$, Melvyn Wright$^2$ \\
\footnotesize{$^1$Department of Astrophysics/IMAPP, Radboud University Nijmegen, P.O. Box 9010, 6500 GL Nijmegen, the Netherlands,$^2$Department of Astronomy and Radio Astronomy Lab, University of California, Berkeley, CA, USA,$^3$Joint ALMA Observatory, ESO, Santiago, Chile,$^4$Academia Sinica Institute of Astronomy and Astrophysics, 645 N. A'ohoku Pl., Hilo, HI 96720, USA,$^5$Max-Planck-Institut f{\"u}r Radioastronomie, Auf dem H{\"u}gel 69, 53121 Bonn, Germany,$^6$Astronomy Department, University of Illinois, 1002 West Green Street, Urbana, IL 61801, USA,$^7$National Radio Astronomy Observatory, 520 Edgemont Road, Charlottesville, USA,$^8$Astronomical Institute 'Anton Pannekoek', University of Amsterdam, Science Park 904, 1098 XH Amsterdam, the Netherlands,$^9$Department of Physics, Astrophysics, University of Oxford, Keble Road, Oxford OX1 3RH, UK,$^{10}$School of Physics and Astronomy, University of Southampton, Highfield, Southampton SO17 1BJ, UK,$^{11}$Argelander-Institut f{\"u}r Astronomie, Universit{\"a}t Bonn, Auf dem H{\"u}gel 71, D-53121 Bonn, Germany}}

\date{\today}
 
\twocolumn[
\begin{@twocolumnfalse}
\maketitle
\begin{abstract}Radio and mm-wavelength observations of Sagittarius A* (Sgr A*), the radio source associated with the supermassive black hole at the center of our Galaxy, show that it behaves as a partially self-absorbed synchrotron-emitting source. The measured size of Sgr A* shows that the mm-wavelength emission comes from a small region and consists of the inner accretion flow and a possible collimated outflow. Existing observations of Sgr A* have revealed a time lag between light curves at 43 GHz and 22 GHz, which is consistent with a rapidly expanding plasma flow and supports the presence of a collimated outflow from the environment of an accreting black hole. Here we wish to measure simultaneous frequency-dependent time lags in the light curves of Sgr A* across a broad frequency range to constrain direction and speed of the radio-emitting plasma in the vicinity of the black hole. Light curves of Sgr A* were taken in May 2012 using ALMA at 100 GHz using the VLA at 48, 39, 37, 27, 25.5, and 19 GHz. As a result of elevation limits and the longitude difference between the stations, the usable overlap in the light curves is approximately four hours. Although Sgr A* was in a relatively quiet phase, the high sensitivity of ALMA and the VLA allowed us to detect and fit maxima of an observed minor flare where flux density varied by $\sim$10\%.
The fitted times of flux density maxima at frequencies from 100 GHz to 19 GHz, as well as a cross-correlation analysis, reveal a simple frequency-dependent time lag relation where maxima at higher frequencies lead those at lower frequencies. Taking the observed size-frequency relation of Sgr A* into account, these time lags suggest a moderately relativistic (lower estimates: 0.5c for two-sided, 0.77c for one-sided) collimated outflow.
\vspace{10pt}
\end{abstract}
\end{@twocolumnfalse}
]


\section{Introduction}

The radio source Sagittarius A* (Sgr A*) at the center of our Galaxy is the best-constrained supermassive black hole candidate found thus far (\citealt{Genzel2010, Falcke2013} for a review). Located at a distance of 8.3$\pm 0.4$ kpc from the solar system, its mass is calculated to be $4.3\pm 0.4\cdot 10^6$ $M_{\odot}$ \citep{Eisenhauer2003, Reid04, Ghez08, Gillessen09, Genzel2010}. For a black hole of this mass, Sgr A* seems to be accreting gas at a very low rate of $\lesssim10^{-7}$ $M_{\odot}$ yr$^{-1}$, as was derived from Faraday rotation measures \citep{Bower2005, Marrone2007}.

The emission from Sgr A* between frequencies of 20 GHz and 230 GHz shows flux density variability of a few tens of percent on hour-long timescales, up to 100\% on month-long timescales, as well as occasional flaring behavior \citep{Dexter2013}. In radio, Sgr A* has an inverted spectrum (i.e., rising flux density with increasing frequency) that peaks at the 'submm bump', around 350 GHz, beyond which the spectrum steeply drops in the infrared regime. The radio emission is thought to originate mostly from partially self-absorbed synchrotron radiation emitted farther out from the black hole, while emission at frequencies corresponding to the submm bump  \citep{Falcke1998} of the Sgr A* spectrum is commonly associated with the optically thin emission closest to the black hole \citep{Falcke1998, Shen2005, Bower2006-2, Doeleman2008}. In the mm regime and at longer wavelengths, the flux density variation is thought to arise from local bulk properties (magnetic field strength, gas density, temperature) of the plasma, while the variability seen in infrared and X-rays is mostly attributed to changes in the population of the high-energy tail of the local electron energy distribution \citep{Ozel2000, Markoff2001, Yuan2003,DoddsEden10, Dibi2013}.\\

While the emission mechanisms for the radio and mm-wavelength emission of Sgr A* are understood fairly well, the identification of the emission with specific flow regions is still a subject of debate. For example, an important question is whether the radio emission is generated in a jet \citep{Falcke93} or in a radiatively ineffcient accretion flow \citep{Narayan1995}. Sgr A* in its flaring state fits neatly onto the fundamental plane of black hole activity \citep{Merloni2003, Falcke2004, Plotkin2012}, and as such it would be expected to feature a jet as other sources on that scaling relation do \citep{Markoff2005}. As yet, no direct detection of a jet has been made for Sgr A* despite the claimed presence of tantalizing jet-like features close to the Galactic center on parsec scales \citep{YZ2012, Li2013}. Any putative jet structure close to the black hole cannot be resolved below observing frequencies of $\sim$100 GHz  because interstellar scatter-broadening blurs our view of the Galactic center at such frequencies, an effect that progressively increases with lower frequency \citep{Lo1981, Langevelde1992, Bower2006, Moscibrodzka2014}. At higher observing frequencies, interstellar scintillation is less of a problem - in the mm-wave regime, existing VLBI networks should be able to directly observe the proposed shadow of the event horizon with mmVLBI \citep{Falcke00,Doeleman2008}.\\

There are other ways in which the nature of the emitting gas flow may be determined, however. Sgr A* exhibits an inverted radio spectrum. Flat or inverted radio spectra are commonly seen in quasars and active galactic nuclei, where the bases of radio jets resolved at high resolution show dominant emission at different radio frequencies as a function of distance from the core, which
is due to optical depth effects \citep{Hada11}, as has also been predicted from theory \citep{BK79, FB95}. The multifrequency spectrum of Sgr A* (from radio to X-ray) in its flaring state looks very much like the spectrum of M81*, which has a weak jet \citep{Bietenholz04}. The emission from an unresolved, compact jet may explain the inverted radio spectrum of Sgr A* \citep{Falcke93, Moscibrodzka2013}.\\

Presence of a jet implies that specific correlations should be detected between light curves at different frequencies. As the peak frequency of radio emission changes with position along the jet axis, we expect variations in flux density at different observing frequencies to exhibit time lags relative to one another as the emitting gas moves out. Previous observations of Sgr A* made with the VLA have indeed suggested the existence of a time lag of $\sim$20  to 40 minutes in flux density variability between light curves measured at 43 GHz and 22 GHz, with variability in the higher-frequency lightcurve leading that in the lower-frequency lightcurve \citep{YZ2006, YZ2008}. Yusef-Zadeh et al. interpreted this as emission from an expanding plasma cloud \citep{vdLaan1966} with velocities reaching about 0.01$c$, but this interpretation does not take VLBA sizes into account. When coupled to the observed relation between the observing frequency and the measured intrinsic size of Sgr A* \citep{Bower04, Doeleman2008}, the time lag between 43 GHz and 22 GHz corresponds to a size difference of $\sim$30 light minutes. Thus, such a time lag suggests the presence of a fast and directed outflow with a moderately relativistic speed \citep{Falcke2009}. As the emission at observing frequencies below the submm bump is probably all partially self-absorbed synchrotron emission, time lags may be present between the light curves at any two different frequencies in that region. Measurements of time lags over a wider range of frequencies are of interest as they may aid in establishing a flow velocity profile, and they may even provide an estimate of how close to the black hole the outflow can be traced.\\

\section{Observations and data reduction}

Our VLA observations were taken on May 18, 2012 from 05:25:15 UT to 12:54:01 UT in CnB configuration, chosen to coincide with a Chandra observation. Light curves for Sgr A* were taken in pairs of subbands for three basebands (X, Ka and Q), yielding 1-GHz-wide subbands at center frequencies of 19 and 25.5 GHz (K-band), 27.48 and 37.99 GHz (Ka-band), and 39.55 and 48.5 GHz (Q-band) - each using 30-second scans at an integration time of 3 seconds. The subbands eventually used for each center frequency were 4-7 (19 GHz), 0-7 (25.5 GHz), 0-7 (27 GHz), 0-7 (37 GHz), 2-7 (39 GHz), and 1-2 (48 GHz). Flux and bandpass calibration were made on the standard VLA calibrator 3C286. A monitoring loop with a period of 7.5 minutes was used for Sgr A*: within each iteration of this loop, J1744-3116 was used as a gain calibrator source and J1745-283 was used as a check-source, cycling through all three basebands in turn. The integration time of 3 seconds was chosen such that the RMS noise per scan was expected to be 1 mJy when all subbands in a baseband were used.\\

The VLA data (project code: 12A-339) were initially reduced using the VLA pipeline version 1.2.0 (rev. 9744) on CASA 4.1.0. After running the VLA pipeline, the sufficiently high flux density of all sources allowed us to perform careful phase self-calibration on them using progressively shorter solution intervals down to one integration length (3 seconds). Some subbands were flagged in this calibration process because their calibration solutions
did not converge. The subbands that remained unflagged were (format: baseband (subbands)) 48 GHz (1,2), 39 GHz (2-7), 37 GHz (all), 27 GHz (all), 25 GHz (all), and 19 GHz (4-7). The declination of Sgr A* combined with the latitude of the VLA means that Sgr A* never reaches an elevation over 27 degrees for the VLA. Therefore the first and last parts (before approximately 6:40 UT and after approximately 11:50 UT) of the observation suffer from coherence loss: the source is less than 18 degrees above the horizon, and the effective path length through the atmosphere for the signal varies rapidly and strongly between antennas. As such, all data in these time windows were flagged before recalibration.\\

For the ALMA track, observations were made in ALMA cycle 0 on May 18, 2012 from 03:30:47 UT to 10:52:16 UT (project code: 2011.0.00887.S). The ALMA light curves for Sgr A* were taken at ALMA bands 3, 6, and 7 using pairs of spectral windows each centered on 95, 105, 247, 260, 338, and 348 GHz. Each pair of spectral windows covered 3.75 GHz bandwidth for a total of 7.5 GHz bandwidth per ALMA band. Scan lengths were chosen to yield a sensitivity of 0.5 mJy. All individual scans of Sgr A* were bracketed by either NRAO530 or J1924-292 as calibrators, while flux density calibration was made on Titan and Neptune. At that time, ALMA had 19 antennas available. In this paper we limit discussion to the 100 GHz ALMA data: the light curves at 250 GHz and 340 GHz will be the subject of a future paper.\\

For ALMA, the source setup was somewhat more complicated because all observations in cycle 0 had to be obtained in two-hour blocks. This means that the ALMA dataset consists of four separate blocks that are contiguous in time, each starting with a flux density calibrator measurement (Titan for the first two blocks, Neptune for the latter two). Within each block, five scans on Sgr A* were made where each of these was bracketed by scans on one of the calibrator sources NRAO530 and J1924-292, a precaution taken because of possible calibration difficulties that might otherwise occur in cycle 0. Thus the scan setup for each block was (using the first letter of each source) 'NSN JSJ NSN JSJ NSN', the average switching time between bracketing calibrator scans and Sgr A* scans was 2 minutes, and the time cadence on Sgr A* in each band was 15 minutes. As J1924-292 exhibited irregular results in its polarization-dependent flux density levels, it was not used as a calibrator and only used as a check-source. Gain levels were stable enough to warrant the usage of NRAO530 as the only gain calibrator. The ALMA data were calibrated using a custom script based on the calibration procedure for the QA2 process of Cycle 0 data, with subsequent phase self-calibration on Sgr A* (which has a very strong unresolved component of around 2.5 Jy at 100 GHz) using baselines longer than 150 k$\lambda$.\\

\begin{figure*}
\hspace{0pt}
\includegraphics[width=0.5\textwidth]{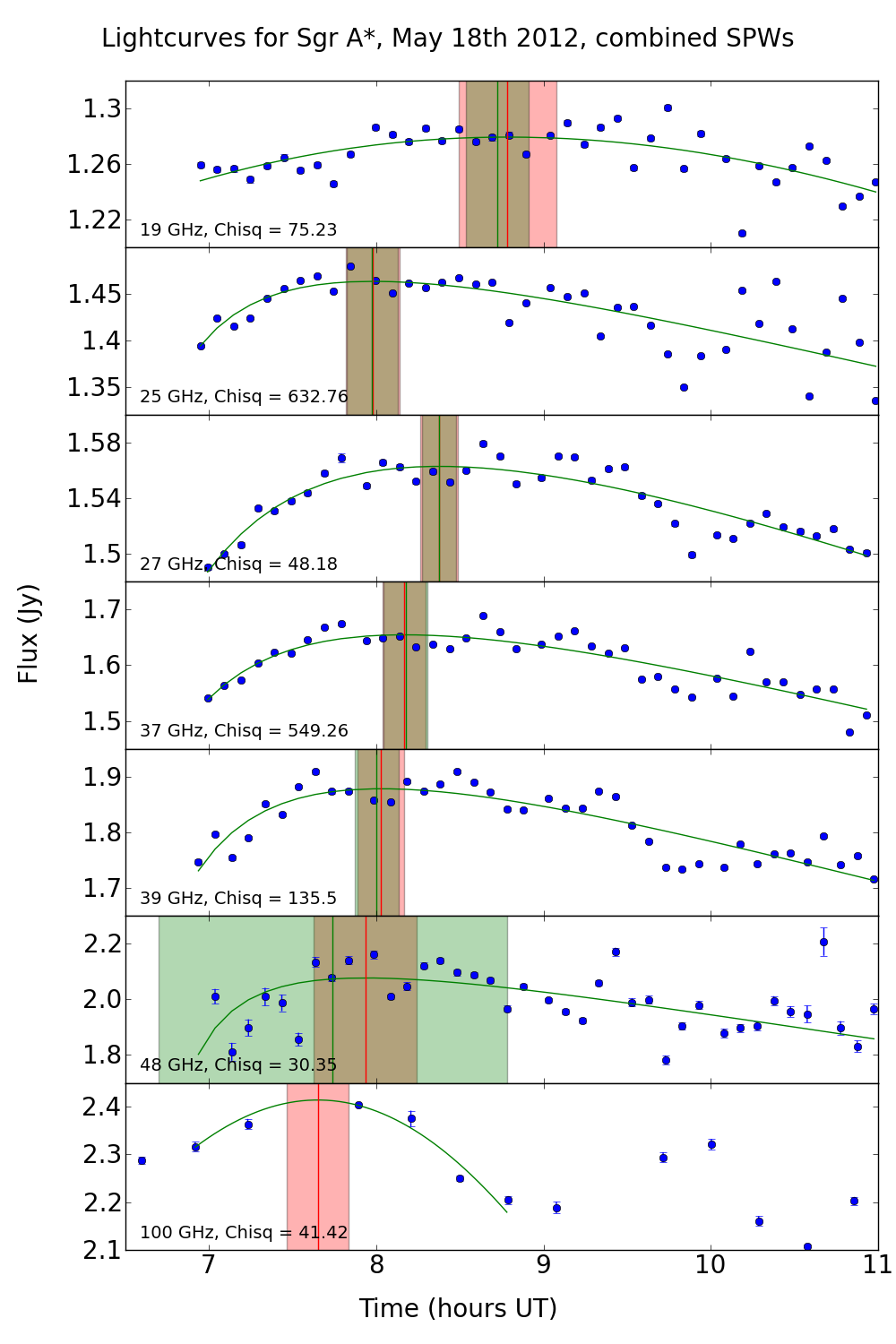}
\includegraphics[width=0.5\textwidth]{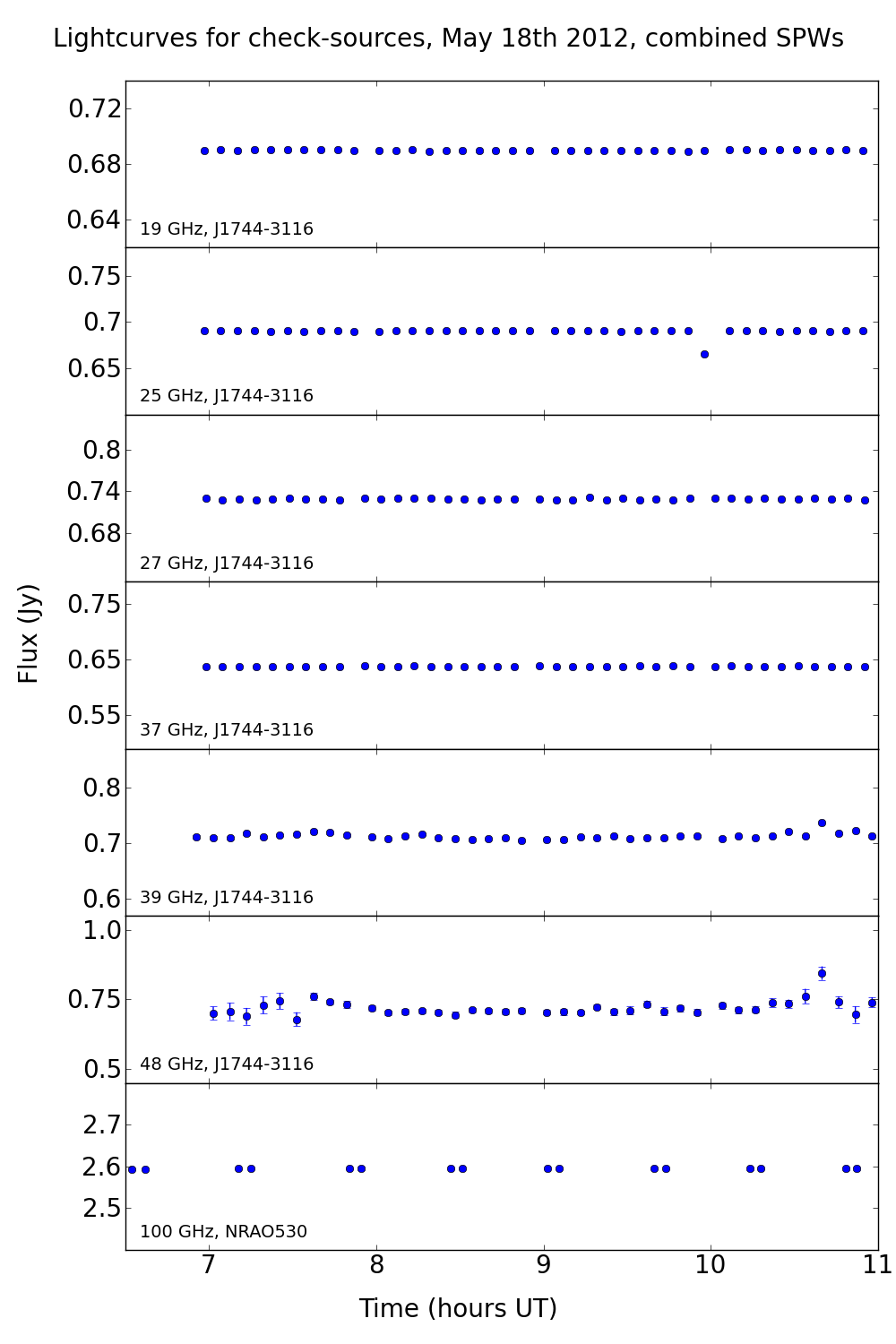}
\caption{{\bf Left:} Flux vs. time of Sgr A* for the VLA data (top six graphs) and the 100 GHz ALMA data (bottom graph). The fitted FRED function is plotted as a green (gray) curve, the position of the maxima is plotted as a vertical red (dark gray) line with the uncertainty in the fit superimposed as a red-shaded (dark gray) region. Green-shaded (light gray) regions indicate the uncertainty on the fits obtained by randomly dropping half of the data points for 500 iterations - see Sect. 3 for details. {\bf Right:} Flux vs. time for the calibrator sources (J1744-3116 for the VLA data, NRAO530 for the ALMA data).}
\label{fig-fits}
\end{figure*}

From the calibrated VLA and ALMA data, the light curves for Sgr A* were obtained by averaging all unflagged UV visibilities per scan from all baselines longer than $150\hspace{3pt}\textrm{k}\lambda$, using uniform weighting for the selected baselines. We chose to only use projected baseline lengths over $150\hspace{3pt}\textrm{k}\lambda$ because we wished to avoid any contamination from the extended emission around Sgr A*, and baselines shorter than the chosen length show hints of structural variations over the track. Limiting ourselves to the longer baselines enabled us to work directly with the visibilities rather than needing the additional steps of imaging and model fitting. The errors on the flux density levels were estimated using the spread in amplitude of the calibrated visibilities over each scan. The calibrated light-curve data can be seen in the left column of Fig. \ref{fig-fits}. All light
curves can be seen to exhibit a gradual rise and decay, with shorter-timescale variation superimposed. For the calibrated VLA data, the resulting noise levels for the basebands at 19, 25, 27, 37, 39, and 48 GHz are 1.4, 0.7, 1.2, 0.9, 2.6, and 12.1 mJy, respectively. Limited calibration accuracy and the flagging of several subbands degrades the sensitivity from the desired sensitivity in the highest frequency bands. The final uncertainties in flux density are dominated by the calibration uncertainty coming from the variability that was exhibited by the check-source: this brings the total relative flux density measurement uncertainty to approximately 5\% for the VLA data. For the ALMA data, the uncertainty in flux density is approximately 5\% as well - however, the errors there are relatively small compared to the intrinsic variability of Sgr A* at 100 GHz. We note that the flux calibration of the VLA data does not give a perfectly smooth spectrum for our calibrator source J1744-3116, but this only affects the overall flux levels of the Sgr A* light curves by $\sim$5\% and does not affect the conclusions in this paper.\\

\begin{figure}
\includegraphics[width=0.5\textwidth]{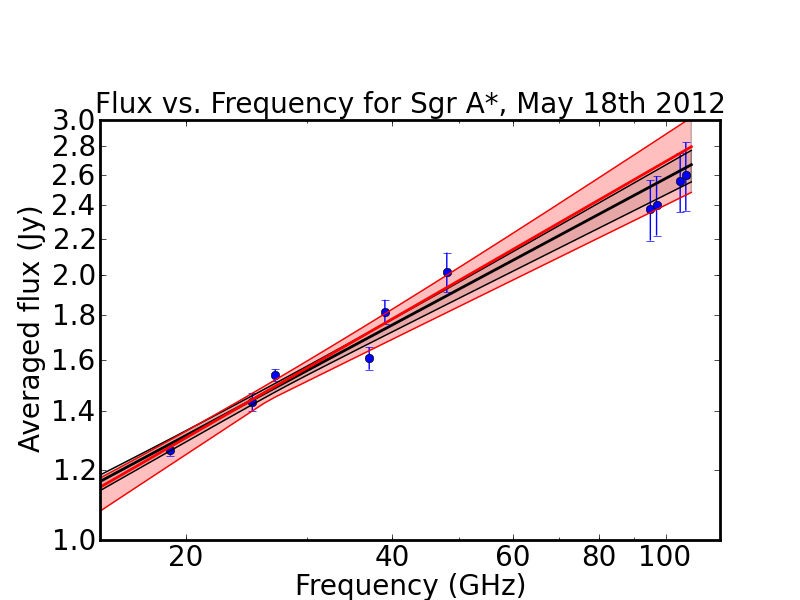}
\caption{Averaged flux density as a function of frequency for all light curves for the time period between 7h and 11h UT. The error bars denote measurement error convolved with flux density variability over the track, where variability is the dominant contribution. Variability is strongest at the highest frequencies. The spectral index obtained from using VLA + ALMA data is indicated by the black line (with 1-sigma fit uncertainties filled in with gray) and yields $\alpha = 0.42 \pm 0.03$. Using the VLA data only (red/gray line, with red/light gray 1-sigma uncertainty region), $\alpha = 0.45 \pm 0.07$.}
\label{fig-fluxfreq}
\end{figure}

\section{Analysis and results}

The spectral energy distribution of Sgr A* across the measured frequencies (see Fig. \ref{fig-fluxfreq}) has a spectral index of $\alpha = 0.41 \pm 0.03$ (with $\alpha$ defined as in $S_\nu \propto \nu^{\alpha}$) when all data (VLA + ALMA) are used for the spectral fit, and we obtain $\alpha = 0.50 \pm 0.07$ when only the VLA data are used. While this difference in spectral index cannot be called significant, it hints at a flattening of the spectrum as the submm bump is approached.\\

The VLA light curves at first sight each show a similar evolution of flux density with time: a rise in flux density level between 7h and 8h/9h UT, and a more slowly diminishing flux density beyond 8h/9h UT. The observed longer-term flux evolution over the full track is overlaid with more rapid variations in measured flux, which occur simultaneously in all frequency bands. These fluctuations are probably caused by atmospheric influence, which causes varying coherence loss as a function of time. The ALMA flux density measurements are highly precise, with a very small spread in visibility values per scan. The time cadence, however, is coarser than it is for the VLA data. Nonetheless, the evolution of flux density with time can be distinguished with high significance. At 100 GHz the flux density evolution is smooth, and there is a local maximum in flux at around 7:45h UT, followed by a later peak around 10:00h UT.

\begin{figure*}
\hspace{0pt}
\includegraphics[width=0.5\textwidth]{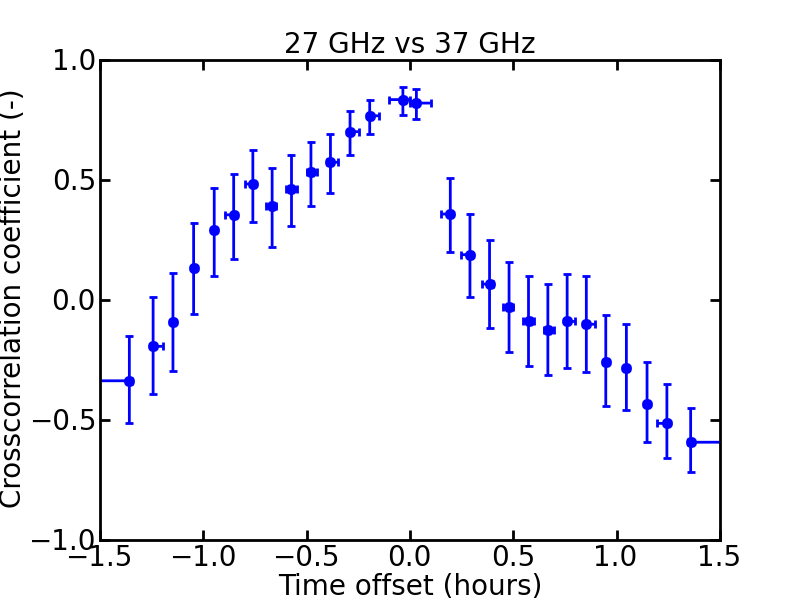}
\includegraphics[width=0.5\textwidth]{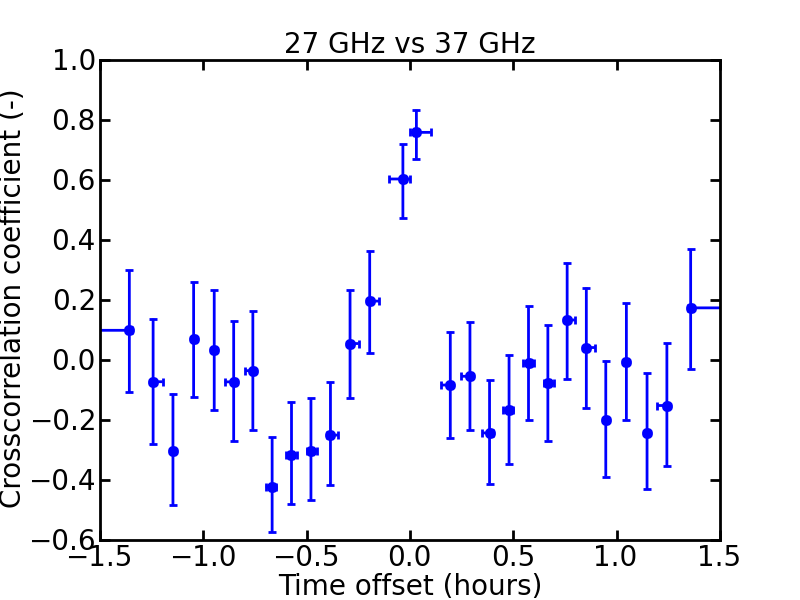}
\caption{Example of a cross-correlation curve for the original light-curve data (left) and for the light-curve data with the FRED trend subtracted (right). The skew in the CCF for the full light curves suggests a nonzero timelag, but the zero-lag peak is too prominent to provide any useful estimate. After the fitted FRED trend is subtracted from both light curves and the cross-correlation is performed again, the only prominent cross-correlation peak corresponds to zero time lag.}
\label{fig-var-cc}
\end{figure*}

The z-transform discrete correlation function (ZDCF) algorithm \citep{Alexander1997} provides a way to cross-correlate light
curves that have uneven temporal sampling. This method for finding time lags between the VLA light curves yields a strong zero-lag component in every case (see Fig. \ref{fig-var-cc} for an example), coming from the short-time fluctuations in the data and probably attributable to coherence loss. These zero-time lag spikes tend to dominate the cross-correlation curves. Although skew is apparent in most cross-correlation curves, the zero-lag peaks preclude any meaningful time lag estimates to be made this way. To derive reliable time lag estimates, we chose the simple and robust approach of fitting the longer-term flux density evolution in all light
curves. To establish the times at which flux density maxima occur in these light curves, we fitted a smooth function to this general trend, allowing for different timescales to be associated with the rise and fall. Based on the general shape of the light curves, the choice was made to employ 'fast rise, exponential decay' (FRED) functions as fitting functions \citep{Bhat1994} (widely used in GRB light curve fitting). These functions consist of the product of two exponentials, involving four free parameters:\\

\begin{equation}
f(t)=A\cdot e^{2\sqrt{b/a}}\cdot e^{-(t-\Delta)/a-b/(t-\Delta)},
\label{FREDfunction}
\end{equation}
where $A$ is the maximum value of $f(t)$, $a$ and $b$ are parameters controlling the slopes of either side of that maximum, and $\Delta$ is the value of $t$ for which the maximum value is reached. Because the FRED flux density value rises up from zero and returns to zero as its argument $t$ is left running, it is not a suitable function to use over time intervals that are too long: we are only interested in using it to fit a local and asymmetric feature in the light curves. To keep the general shape of the light curves sufficiently simple to enable the fit without sacrificing too many data points, we only used the flux density measurements directly around the bump feature (7h - 11h UT for VLA, 6:50 - 8:50 UT for ALMA).  Acting on the assumption that this feature in the ALMA light curve can be attributed to the same event in the source that caused the maxima found in the VLA light curves, we used the fitting algorithm on this feature as well.\\

Variations on shorter timescales in the VLA light curves may affect the fit results. Therefore we also performed the fits using a Monte Carlo approach in which, at random, half of the data points (20 out of 40) for a given VLA light curve were dropped before attempting a fit with the FRED function. For each light
curve, this was iterated 500 times. The resulting times of maxima for all obtained fits were averaged, and the standard deviation of their distribution was calculated. These results are also plotted in Fig. \ref{fig-fits}, shaded in green. The large standard deviation in the 48 GHz case indicates skew in the range of predictions made by the Monte Carlo method, causing the left limit to fall outside of the data range. High reduced $\chi$-squared values indicate that we fitted light curves whose evolution is more complicated than can be grasped by a simple function; significant short-term variability remains from imperfect flux calibration. The absolute flux density uncertainties are dominated by the calibration uncertainty of $\sim$5 percent for the entire light
curve (not shown in the figure). For most light curves, these results correspond well to the fit for which all data were used. For the light curve at 48 GHz the fits are not as robust. This is most probably due to the relatively strong flux density variations at short timescales for the 48 GHz light curve, combined with the short FRED rise time. The resulting best-fit values and uncertainties for the times of flux density maximum are shown in Fig. \ref{fig-fits}, and the time lags we found are summarized in Table \ref{table-timelags} where the middle column uses all data and the rightmost column uses the average result from the Monte Carlo approach. As the ALMA data does not have many measurements within the relevant time window, the Monte Carlo approach could not be used there.\\

\begin{table}
\centering
\caption{Times of flux density maxima from curve-fitting of individual light curves. Column 1: all data, Col. 2: subsample with MC}
\label{table-timelags}
\begin{tabular}{rll}
Frequency & Time of max (hrs UT) (1) & Time of max (hrs UT) (2)\\
\hline\\
100 GHz & $7.65\pm0.19$ & - \\
48 GHz   & $7.93\pm0.31$ & $7.82\pm0.90$ \\
39 GHz   & $8.03\pm0.14$ & $8.01\pm0.12$ \\
37 GHz   & $8.17\pm0.13$ & $8.17\pm0.13$ \\
27 GHz   & $8.38\pm0.11$ & $8.38\pm0.11$ \\
25 GHz   & $7.98\pm0.16$ & $7.97\pm0.15$ \\
19 GHz   & $8.78\pm0.29$ & $8.72\pm0.19$ \\
\end{tabular}
\end{table}

\begin{figure}
\includegraphics[width=0.5\textwidth]{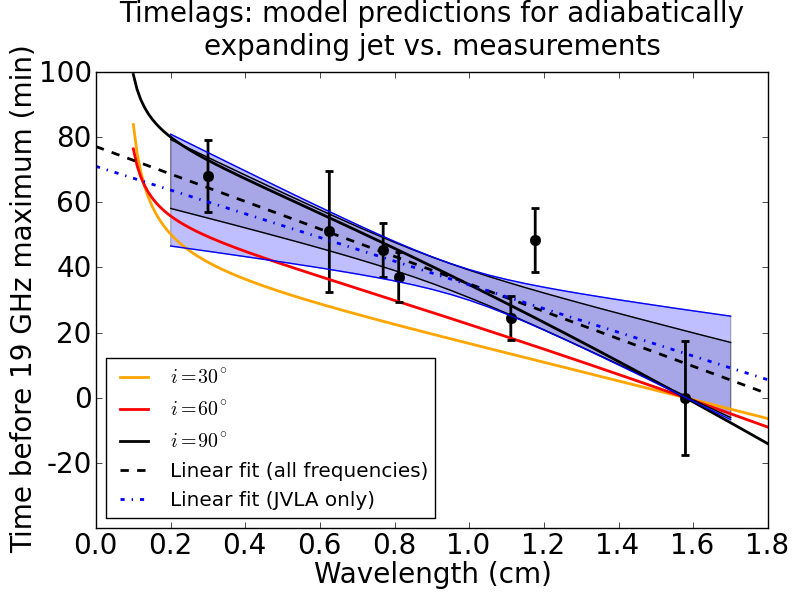}
\caption{Fitted times for the flux density maxima in each baseband (relative to the 19 GHz (1.6 cm) maximum), plotted as a function of observing wavelength. The figure uses the errors on the fit maxima obtained from fitting all data points. Dark shaded regions indicate the uncertainty in fit slope using all light curves, lighter shaded regions indicate the uncertainty in fit slope from VLA light curves alone. The slopes obtained are $42 \pm 14$ mins/cm (all data) and $36 \pm 21$ mins/cm (VLA data alone). The continuous lines are time-lag predictions from the jet model by \citet{Falcke2009} assuming inclinations of 30, 60, and 90$^\circ$.}
\label{fig-lags}
\end{figure}

\begin{figure}
\hspace{0pt}
\includegraphics[width=0.5\textwidth]{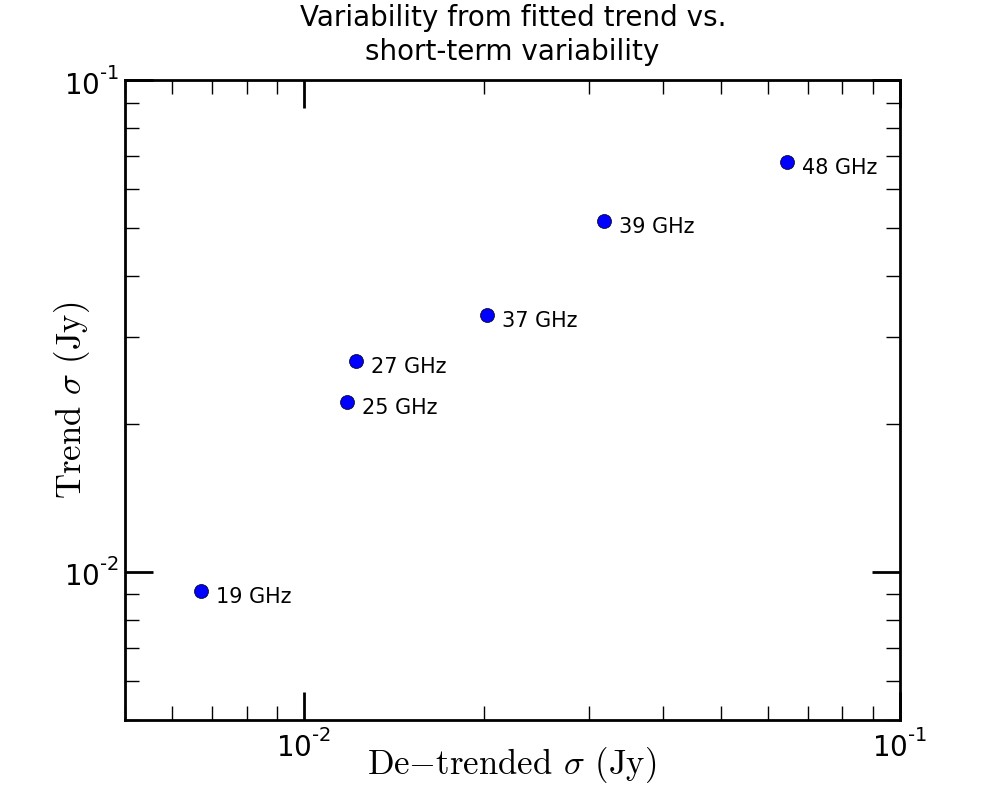}
\caption{Variability due to FRED-fitted light-curve trends (vertical axis) versus remaining variability in trend-subtracted light
curves (horizontal axis) for the VLA data. Both axes show standard deviations. These quantities indicate that the long-term variability that is fitted by the FRED function scales with the remaining short-term variability.}
\label{fig-var-trend-res}
\end{figure}

With the 25 GHz data as notable exception, the maxima occur at later times for the lower observing frequencies, which is compatible with a linear relation between observing wavelength and time of maximum flux density. There is always the risk of misinterpreting the ALMA bump at 7:40h UT as being causally connected to the maxima in flux density found in the VLA data. To check whether the VLA data by themselves are  consistent with the existence of these time lags, a separate fit was made using only the VLA data (Fig. \ref{fig-lags}), and the fit results are practically identical to those obtained with the full data set. The time lag expected for the ALMA peak based on the VLA time lag/frequency fit coincides with the measured value, and thus the identification of the ALMA flux density maximum as being related to the VLA flux density maxima seems justified. The fact that the maxima occur at different times for different frequencies also precludes interpreting the observed flux density evolution as being a purely atmospheric or elevation-dependent effect. Any elevation-dependent change in measured flux density would impose a simultaneous rising and falling of all light curves, which is not what we observe.\\

The FRED fits follow the general, long-term trend that is present in the data. In addition to this general trend, all light curves exhibit shorter-timescale fluctuations. Cross-correlation analysis on the original light curves therefore shows correlation contributions from both the general trends in the data and the shorter-timescale fluctuations. If these two variability components do not exhibit the same time lag between frequencies, interpreting the cross-correlation curves is problematic. To deal with this problem, we have subtracted the fitted FRED trends from all light curves and performed a cross-correlation analysis on the de-trended light curves (see Fig. \ref{fig-var-cc}). This cross-correlation peaks at zero time lag, suggesting that the short-term fluctuations have a different origin from the long-term trends. The most likely cause of the short-term fluctuations are calibration residuals stemming from phase-coherence loss due to the low elevation of Sgr A* at the VLA site. While there is a correlation between the FRED variability and the residual variability as seen from Fig. \ref{fig-var-trend-res}, this does not imply that they share the same cause. Atmospheric influence is stronger for higher observing frequency, and this effect is unrelated to intrinsic source variability.

\section{Discussion and conclusion}

The time lags across this broad range of frequencies corroborate the picture of an expanding plasma flow with a diminishing optical depth over time. When the time lags found in this work are combined with the existing results from \citet{YZ2006}, they are compatible: Yusef-Zadeh et al. reported a time lag between 43 GHz and 22 GHz of 20 to 40 minutes, while we detect a time lag of 28 $\pm$ 9 minutes between these two frequencies when we use the linear time lag/wavelength fit based on our measurements (see Fig. \ref{fig-lags} and caption).\\

Measurements on time lags between Sgr A* light curves at 102 and 90 GHz were performed by \citet{Miyazaki2013}, and they reported the time delay between 102 GHz and 90 GHz as being $-2.56 \pm 0.9$ min (i.e., the 90 GHz light curve is leading the 102 GHz light curve). The expected time lag between 102 and 90 GHz that would agree with the 43 GHz -22 GHz lag found by \citet{YZ2006} is quoted as being close to 3 min (with the precise value depending on the index of the power-law distribution in electron energy), whereas an extension of the linear relation we find in this work predicts a time lag between these frequencies of $1.7 \pm 0.6$ min (this figure increases somewhat if low inclination angles are considered, see Fig. \ref{fig-lags}). We stress that the models used to predict the time lags are very simple in all cases, and in particular measured time lags between closely neighboring frequencies can deviate from the predicted relation due to more complex plasma flow properties close to the black hole. The Blandford-K\"onigl jet model uses a $\tau=1$ surface, the location of which along the jet only depends on the accretion rate, and which is constant throughout the jet cross-section. The actual nature of any outflow may locally be of a more chaotic character, with different regions in the jet cross-section having different plasma densities and different optical depths, as is typically witnessed in GRMHD simulations \citep{Moscibrodzka2014}. Our measurements, obtained over a broad range of frequencies, are expected to sample a greater spatial range of the proposed outflow. They should hence provide a robust result and a characterization of the behavior of the system as a whole, and we believe this approach warrants the use of a simple outflow model.\\

\begin{figure}
\includegraphics[width=0.5\textwidth]{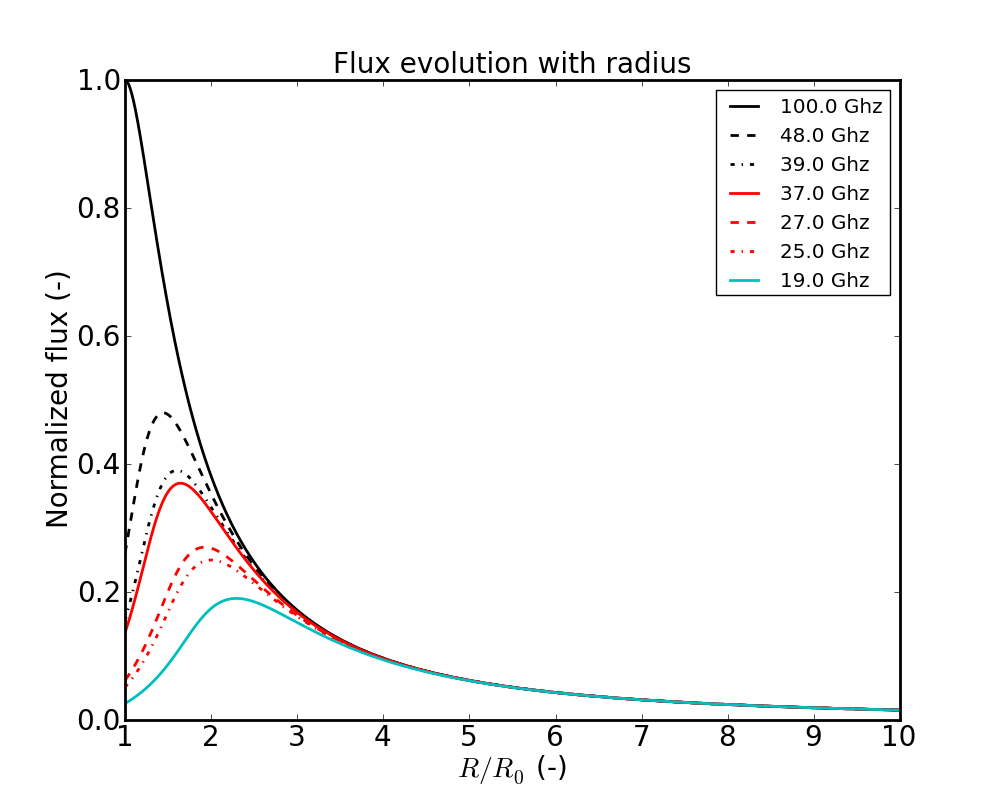}
\caption{Normalized flux density profiles of an adiabatically expanding plasma cloud as a function of radius for all frequencies. The continuous flux profiles use the expressions given in \citet{YZ2008}, with a particle spectral index of 1 as was determined to be the best-fitting value in that work. Note that the initial radius used by \citet{YZ2008} ($\sim$3 Schwarzschild radii) is different from the initial radius that follows from using the size-frequency relation ($\sim$ 11 Schwarzschild radii).}
\label{YZflareprofile-radius}
\end{figure}

\begin{figure}
\includegraphics[width=0.5\textwidth]{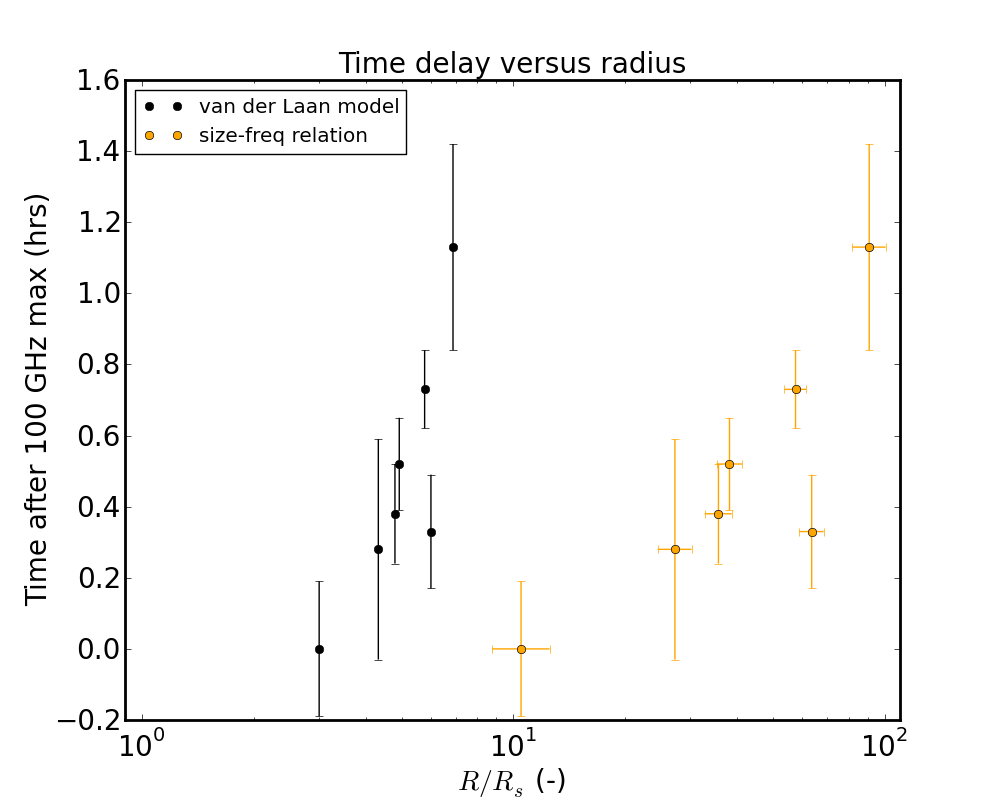}
\caption{Times of flux density maxima for different frequencies plotted against the radii at which maximum flux density is reached. Black data points are obtained by combining our time lags with the outflow model used in \citet{YZ2008}, see also Fig. \ref{YZflareprofile-radius}. If the initial radius is a only few Schwarzschild radii, the estimated flow velocity reached by the expanding plasma is on the order of 0.03$c$. Orange (light gray) data points are obtained by combining our time lags with the size-frequency relation (Eq. \ref{angularsize}), and yield a velocity of $\sim$0.5$c$. Note that the initial radii are not the same for both cases.}
\label{YZflare-evolution}
\end{figure}

Without source size measurements, many different models of expanding plasma flows can be made to fit our observations. Following the analysis by \citet{YZ2008}, where an adiabatically expanding plasma cloud was used as a model for flare occurrence in Sgr A*, the cloud expands to just $\sim$2.3 times the size it initially has at the 100 GHz maximum (see Figs. \ref{YZflareprofile-radius} and \ref{YZflare-evolution}) for the frequency range we record. The radii at which the lower-frequency emission peaks is only a few times the initial radius (which is taken to be 3 Schwarzschild radii). If we adopt the initial radius of the cloud as being $\sim$3 Schwarzschild radii, as was done by \citet{YZ2008}, the associated flow velocities that occur according to this model are only around 3 percent of the speed of light. This is twice the velocity that was found by \citet{YZ2008}.\\

However, we can use the size-frequency relation of Sgr A* as presented in \citet{Falcke09} as additional information with which to provide a general estimate of the gas flow velocity. Although we do not have sufficient information to identify the variable emission (that we focus on) with the quiescent emission for which the size-frequency relation holds, considering the two components to reflect the same gas flow is the simplest hypothesis that we can consider. The size-frequency relation describes the measured intrinsic (i.e., corrected for interstellar scattering) size of Sgr A* as a function of observing wavelength, and has the form\\

\begin{equation}
\phi_{\textrm{Sgr A*}}=(0.52 \pm 0.03) \textrm{mas} \times (\lambda/\textrm{cm})^{1.3 \pm 0.1}.
\label{angularsize}
\end{equation}

In this expression, $\phi_{\textrm{Sgr A*}}$ is the angular size of Sgr A* on the sky and $\lambda$ is the observing wavelength. Combining this (angular) size-frequency relation with an estimate for the distance between Earth and the Galactic center and taking the difference in source size for two observing frequencies, we obtain an expression for the projected source size difference on the sky in length units. When we assume a source inclination of 90$^\circ$ and a one-sided outflow interpretation (as was done by \citet{Falcke2009}), we find that our data suggest an outflow velocity of $\sim$0.77$c$ ($\gamma\beta\approx1.2$). Using the variability seen in our measurements (Fig. \ref{fig-var-trend-res}) as a proxy for flare amplitude with the size-frequency relation (Eq. \ref{angularsize}), we obtain the data points that are plotted separately in Fig. \ref{YZflareprofile-radius}.\\

\begin{figure}
\includegraphics[width=0.5\textwidth]{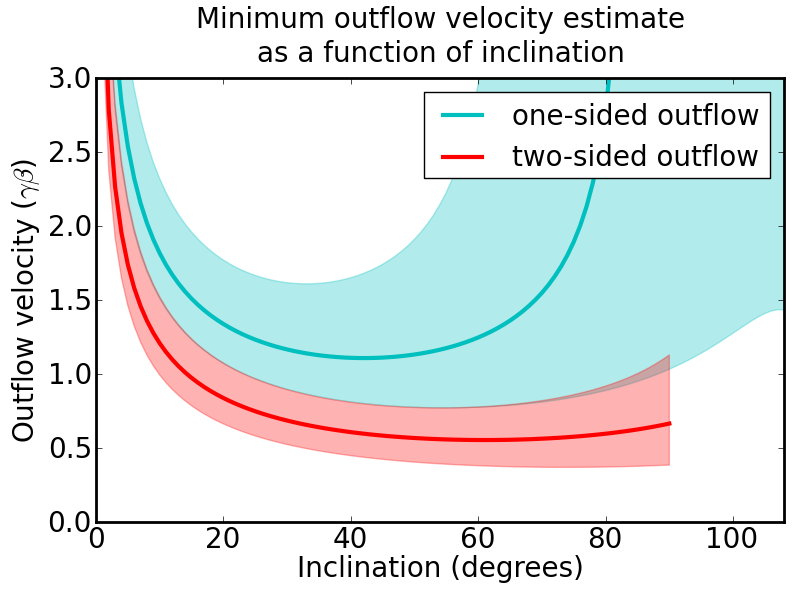}
\caption{Estimates of minimum outflow velocity (expressed in $\gamma\beta$) as a function of the inclination angle of the flow direction. The uncertainties have been calculated from the uncertainty in Galactic center distance (0.4 kpc), the uncertainties in the size-frequency relation (see Eq. \ref{angularsize}) and the uncertainty in time lag vs observing wavelength (14 mins/cm, see Fig. \ref{fig-lags}) using standard error propagation.}
\label{fig-outflowvel-vs-inc}
\end{figure}

Because the source centroid position on the sky as a function of observing frequency is not known, however,we can derive a lower limit on the outflow velocity by assuming identical centroid positions for all observing frequencies. This assumption corresponds to a two-sided jet interpretation (the source grows symmetrically on the sky with lower observing frequency), so for the distance traveled by the gas in one jet we can take half of the intrinsic source size difference. Taking into account the influence of light travel time with different inclination angles yields the following expression for the expected time lag between observing frequencies:\\

\begin{equation}
\Delta_{\textrm{diff}} = \frac{R_{\textrm{SgrA,diff}}}{\sin{i}} \left( \frac{1}{v_f} - \frac{\cos{i}}{c}\right),
\label{lagfreq}
\end{equation}
with $\Delta_{\textrm{diff}}$ the time lag between two observing frequencies, $R_{\textrm{SgrA,diff}}$ the radius difference for the two observing frequencies as calculated using Eq. \ref{angularsize} and our distance to the Galactic center of 8.3 kpc, $i$ the inclination angle (angle between the flow vector and the line of sight from Sgr A* to Earth) and $v_f$ the flow velocity. We can express this relation in terms of $v_f$ and combine it with the relation between time lag and wavelength (see Fig. \ref{fig-lags}). In this way, we can plot the relation between flow velocity estimate and jet inclination angle, where we place the constraint that the flow velocity needs to be positive and lower than the speed of light. Figure \ref{fig-outflowvel-vs-inc} shows this dependence and indicates a minimum flow velocity of $v_f = 0.5$c ($\gamma\beta=0.58$) for an inclination of 60$^\circ$. Including the light travel time in the calculation breaks the symmetry around an inclination of 90 degrees that would otherwise be present. For the two-sided outflow, we assumed here that the outflow component with an inclination smaller than 90 degrees is the one that is picked up in the time lag measurements. For the two-sided jet interpretation only inclinations close to 90$^\circ$ can be modeled reasonably in this way. Inclinations deviating significantly from 90$^\circ$ would result in ambiguous time lags because of the different light travel times from the gas in the two jets. For the case of a one-sided jet, where the source only grows toward one side with lower observing frequency, the lower velocity bound is approximately $v_f = 0.77$c ($\gamma\beta=1.2$). Figure \ref{fig-lags} plots the model-predicted time lags for different inclination angles and shows that no strong constraints are posed on the inclination angle by these data.\\

As these results are based upon measurements of a light-curve feature from a single track, it is prudent to perform this analysis on more light curves as they become available to verify the picture we establish. Although the broad nature of the flux density feature that we used in the FRED fit generally agrees with the result of the ZDCF analysis, light curves with significant flux density changes over shorter timescales would offer an opportunity of using the ZDCF more effectively as an alternative verification of the FRED fitting results. ALMA observations at a higher time cadence taken contemporaneously with VLA observations would facilitate the cross-matching of light-curve features between these frequencies.\\

To summarize, we have measured time lags in Sgr A* light curves from 100 GHz to 19 GHz using ALMA and the VLA. Higher-frequency light curves are seen to have their maxima at earlier times than the lower-frequency light curves. Coupled to the size-frequency relation for Sgr A*, these measurements indicate a moderately relativistic, directed outflow from Sgr A*.

\section{Acknowledgements}
This work was performed as part of the Chandra XVP collaboration (see http://www.sgra-star.com/collaboration-members/ for a list of collaboration members). This paper makes use of the following ALMA data: ADS/JAO.ALMA\#2011.0.00887.S. ALMA is a partnership of ESO (representing its member states), NSF (USA) and NINS (Japan), together with NRC (Canada) and NSC and ASIAA (Taiwan), in cooperation with the Republic of Chile. The Joint ALMA Observatory is operated by ESO, AUI/NRAO and NAOJ. We thank our referee for providing useful comments that improved the presentation of this paper. Reproduced with permission from Astronomy \& Astrophysics, \textcopyright ESO.

\begin{figure}
\includegraphics[width=0.5\textwidth]{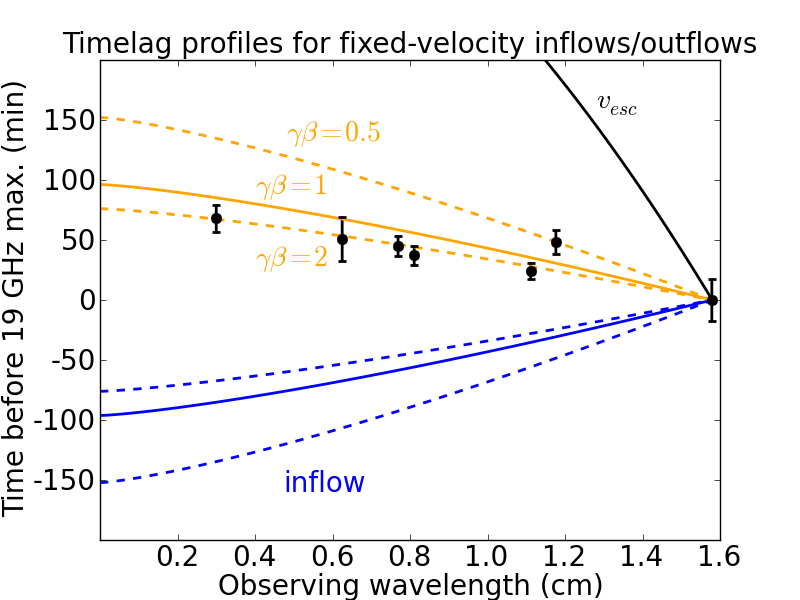}
\caption{Time delays as a function of frequency, expressed in minutes before the time of maximum flux density at $\lambda=1.57$ cm (19 GHz). Colored lines indicate the model time-lag predictions for different constant outflow and inflow velocities. Orange (light gray) lines indicate outflow, blue (dark gray) lines indicate inflow. The data points, derived from the full-data FRED curve fits, suggest an outflow with a moderately relativistic velocity, clustering around $\gamma \beta = 2$. This figure is valid for an inclination angle of 90 degrees and is adapted from \citet{Falcke2009}.}
\label{fig-mrel}
\end{figure}

\bibliography{timelag}
\bibliographystyle{apalike}

\end{document}